\documentclass[a4paper]{article}

\usepackage{INTERSPEECH2021}
\usepackage{multirow}

\title{The ByteDance Speaker Diarization System for the VoxCeleb Speaker Recognition Challenge 2021}
\name{Keke Wang, Xudong Mao, Hao Wu, Chen Ding, Chuxiang Shang, Rui Xia, Yuxuan Wang}
\address{\textbf{S}peech, \textbf{A}udio \& \textbf{M}usic \textbf{I}ntelligence (SAMI), ByteDance}
\email{\{wangkeke.362, maoxudong.1994, wuhao.9528, dingchen.2101, shangchuxiang.19, \\rui.xia, wangyuxuan.11\}@bytedance.com}

\begin{document}

\maketitle
\begin{abstract}
This paper describes the ByteDance speaker diarization system for the fourth track of the VoxCeleb Speaker Recognition Challenge 2021 (VoxSRC-21). The VoxSRC-21 provides both the dev set and test set of VoxConverse for use in validation and a standalone test set for evaluation. We first collect the duration and signal-to-noise ratio (SNR) of all audio and find that the distribution of the VoxConverse's test set and the VoxSRC-21's test set is more closer. Our system consists of voice active detection (VAD), speaker embedding extraction, spectral clustering followed by a re-clustering step based on agglomerative hierarchical clustering (AHC) and overlapped speech detection and handling. Finally, we integrate systems with different time scales using DOVER-Lap. Our best system achieves 5.15\% of the diarization error rate (DER) on evaluation set, ranking the second at the diarization track of the challenge.
\end{abstract}

\section{Introduction}
The goal of VoxSRC-21 is to probe how well current methods can recognize speakers from speech obtained 'in the wild'. The data is obtained from YouTube videos of celebrity interviews, as well as news shows, talk shows, and debates - consisting of audio from both professionally edited videos as well as more casual conversational audio in which background noise, laughter, and other artefacts are observed in a range of recording environments. This paper mainly focuses on the fourth track, which is a speaker diarization track, where the task is to break up multi-speaker audio into homogenous single speaker segments, effectively solving 'who spoke when'. 

The VoxSRC-21 provides VoxConverse \cite{chung2020spot} as validation. The VoxConverse is an audio-visual diarization dataset consisting of over 50 hours of multi-speaker clips of human speech, extracted from YouTube videos. A standalone test set without any additional annotation is used for evaluation. According the results from VoxSRC-20\cite{nagrani2020voxsrc}, the DER of dev and test set in VoxConverse varies greatly. Therefore, we first collect the duration and SNR of dev and test set of VoxConverse and test set of VoxSRC-21, separately. We find that the distribution of the VoxConverse's test set and the VoxSRC-21's test set is more closer. And because the VAD and overlapped speech detection tools we apply in system, use part of the dev data in VoxConverse to train the model. So we choose the test set of VoxConverse as primary validation dataset to select modules and systems.

\begin{figure}[t]
  \centering
  \includegraphics[width=\linewidth]{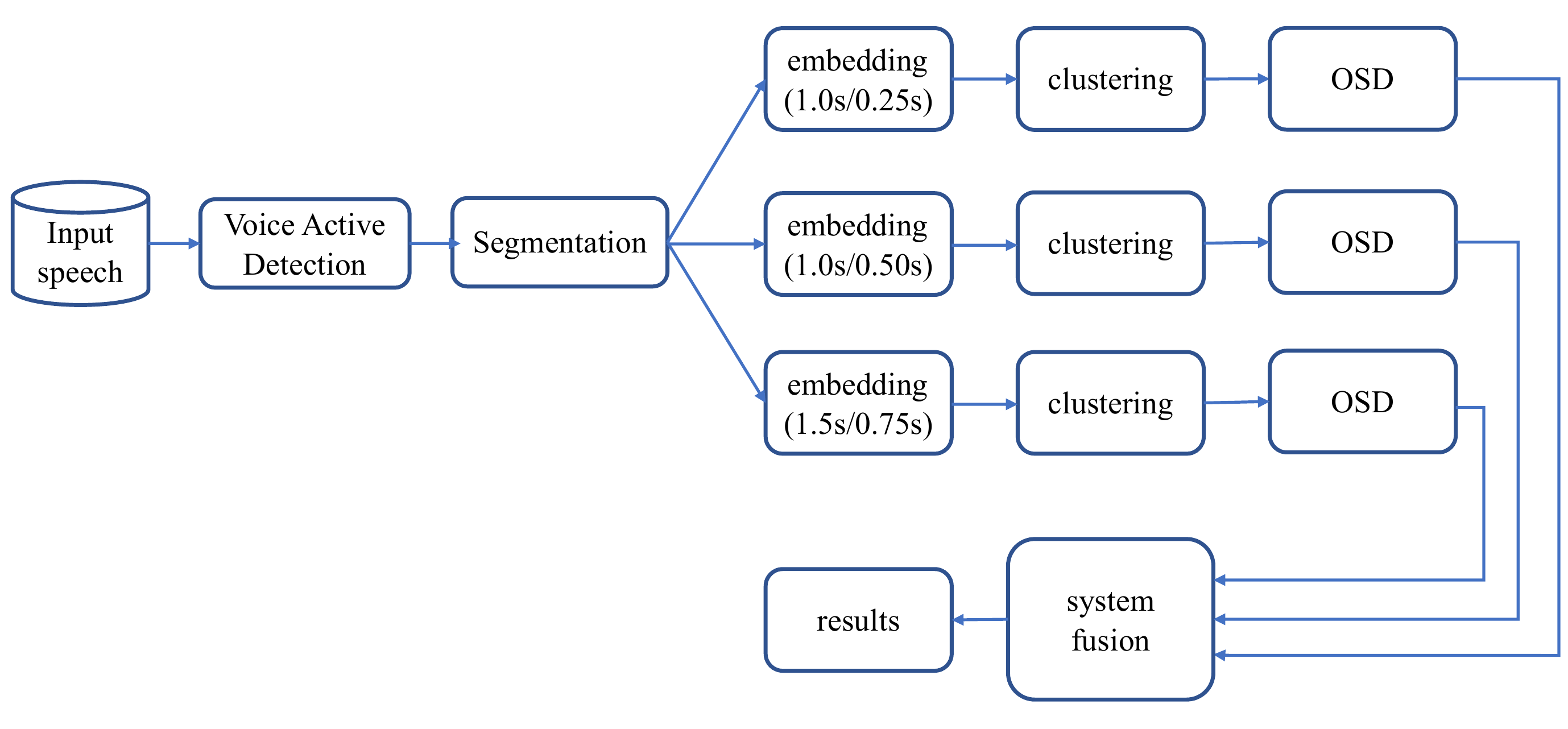}
  \caption{System overview}
  \label{fig:system_overview}
\end{figure}

We propose a clustering-based speaker diarization system which consists of VAD, speaker embedding extraction, clustering, overlapped speech detection and handling, and system fusion shown as Figure~\ref{fig:system_overview}. The prominent components of our system con be summarized as follows.

\textbf{VAD}

We evaluate different VAD systems with label obtained from validation set. VAD of pyannote 2.0 performs the best.

\textbf{Speaker embedding extractor}

An ECAPA-TDNN model based speaker embedding extractor is trained using SpeechBrain toolkit with more data augmentation modules. We evaluate the extractor with VoxCeleb1-test and validation set of VoxSRC-21.

\textbf{Clustering}

We first perform spectral clustering on audio segments with short uniform segmentation, then an re-clustering step is conducted with speaker embedding extracted with longer audio clips.

\textbf{Overlapped speech detection and handling (OSD)}

We use pyannote 2.0 to detect the overlapped speech and assume that there are at most two human voices in overlapped speech. For each overlapped speech segments, we find the two closest speakers in time.

\textbf{System fusion}

We find that system with finer time scale performs better on DER but less robust on clustering, and coarser time scale performs conversely. Thus, we integrate systems with different time scales using DOVER-Lap.

Next section describes the statistics of duration and SNR on the dataset provided by VoxSRC-21. Sections 3 presents the proposed system detailed. Section 4 reports the results and analysis of different systems and clncludes the paper on the last section.

\section{Dataset}
At VoxSRC-21, the development set consists of the development set and test set of VoxConverse (for convenience, in the following article, they are referred to as dev1 and dev2 respectively, and the evaluation set is referred to as test). 

\begin{figure}[t]
  \centering
  \includegraphics[width=\linewidth]{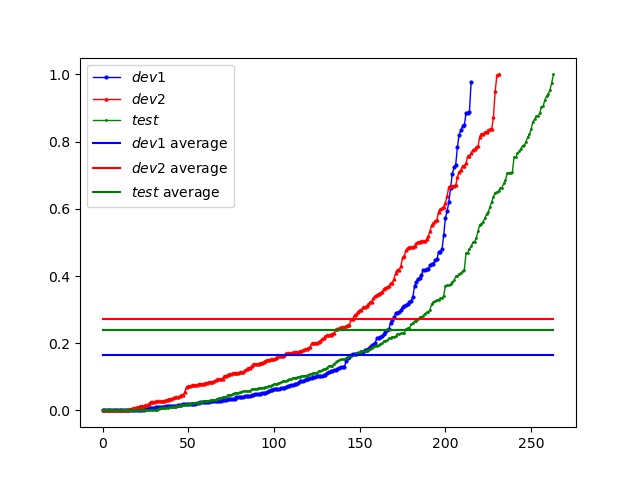}
  \caption{The ratio of low SNR segments.}
  \label{fig:snr_stat}
\end{figure}

A signal noise ratio (SNR) estimation process was firstly applied to dev1, dev2 and test. Each record was divided into 3s segments. The result indicated that the average ratio of low SNR segments without overlap. Then SNR estimation was applied to each segment, followed by counting the proportion of low SNR (\textless5dB) segments. The result indicated that there was a significant gap on the average ratio of low SNR segments between dev1 and dev2, and the condition of test was more similar to dev2, as shown in Figure~\ref{fig:snr_stat}. In addition, we used part of the dev1 data for training in the subsequent vad and overlap detection, so dev2 was mainly used to evaluate the performance of the system.

To alleviate the impact of noise, we tried applying a speech enhancement system as the front-end processing of the speaker diarization. However, the results on dev1 and dev2 showed that although the SNR of the enhanced speech is improved, it also brought damage to speaker voice, which directly led to the decrease of the DER. Therefore, the speech enhancement module was removed from the final system. 

We also made statistics on audio duration, speech duration and number of speakers, as well as overlap proportion. Since reference information was only provided for dev1 and dev2, we only made statistics on these two. As shown in Table~\ref{tab:table1} and Table~\ref{tab:table2}, the average duration of audio in dev2 is longer than that in dev1, as well as more speakers on average, which also intimates that dev2 is more challenging than dev1.

\begin{table}[th]
  \caption{Audio duration and speech duration statistics on dev1 and dev2. Audios refers to the number of records. The format of audio duration and speech duration is minimum/average/maximum duration in seconds.}
  \label{tab:table1}
  \centering
    \begin{tabular}{ c c c c }
    \toprule
    \multicolumn{1}{c}{\textbf{Dataset}} & \multicolumn{1}{c}{\textbf{Audios}} & \multicolumn{1}{c}{\textbf{Audio duration}} & \multicolumn{1}{c}{\textbf{Speech duration}} \\
    \midrule
    dev1  & $216$ & $22/338/1097$ & $4/315/1055$    \\
    dev2  & $232$ & $26/676/1200$ & $25/605/1186$    \\
    dev1+dev2 & $448$ & $22/513/1200$ & $4/465/1186$    \\
    \bottomrule
  \end{tabular}
  
\end{table}

\begin{table}[th]
  \caption{Overlap duration statistics on dev1 and dev2. The format of speakers and overlap ratio is minimum/average/maximum.}
  \label{tab:table2}
  \centering
    \begin{tabular}{ c c c c }
    \toprule
    \multicolumn{1}{c}{\textbf{Dataset}} & \multicolumn{1}{c}{\textbf{Total time(m)}} & \multicolumn{1}{c}{\textbf{Speakers}} & \multicolumn{1}{c}{\textbf{Overlap ratio}} \\
    \midrule
    dev1  & $42.4$ & $1/4.5/20$ & $0\%/3.7\%/36.6\%$    \\
    dev2  & $71.4$ & $1/6.5/21$ & $0\%/3.1\%/38.7\%$    \\
    dev1+dev2 & $113.8$ & $1/5.6/21$ & $0\%/3.3\%/38.7\%$    \\
    \bottomrule
  \end{tabular}
  
\end{table}

\section{System Overview}

\subsection{Overview}

The proposed speaker diarization system is illustrated in Figure~\ref{fig:system_overview}. The input audio is first processed by VAD to obtain valid speech segments. Then uniform segmentation with different time scale is performed on all segments. For each time scale, speaker embeddings are extracted with an ECAPA-TDNN model. Clustering and OSD are conducted separately. Finally, diarization outputs from different time scales are fused by DOVER-Lap. The details of each module will be explained in the following subsections.

\subsection{Voice activity detection}

VAD is the very first step of diarization system and it has a great impact on the final DER result. Though no ground truth VAD labels are provided in this challenge, we can generate VAD labels from diarization labels and evaluate our VAD system on develop set. In order to find the most appropriate vad system, we tried a lot methods and chose \textit{pyannote.audio 2.0}\footnote{\noindent https://github.com/pyannote/pyannote-audio/tree/develop}\cite{bredin2020pyannote} for the following results. We use the model and hyper-parameters hosted in \textit{huggingface}\footnote{\noindent https://huggingface.co/pyannote/segmentation}, which is trained and tuned on a dataset including VoxConverse dev1. We also have tuned the hyper-parameters on \textit{dev1+dev2} and got a little lower VAD error. However, the DER doesn't get better. So we didn't use the tuned hyper-parameters.

Specifically, pyannote has a post-processing step, which fills gaps shorter than a given threshold and removes active regions shorter than another given threshold. 
We found that this post-processing step is also effective for other VAD systems. Because VoxConverse dev1 is already included in pyannote training data, we evaluate VAD results on dev1 and \textit{dev1+dev2}.

\begin{table}[th]
  \caption{Frame level results for different VADs on dev2. VAD1-5 refers to the VAD systems we have tried. pa refers to pyannote. pa\_tuned refers to pyannote with hyper-parameters tuned on development set}
  \label{tab:vad1}
  \centering
    \begin{tabular}{ l c c c c c c }
    \toprule
    \multicolumn{1}{c}{\textbf{System}} & \multicolumn{3}{c}{\textbf{dev1+dev2}} & \multicolumn{3}{c}{\textbf{dev2}} \\
    \midrule
    \multicolumn{1}{c}{\textbf{}} & \multicolumn{1}{c}{\textbf{Miss}} & \multicolumn{1}{c}{\textbf{FA}} & \multicolumn{1}{c}{\textbf{Error}} & \multicolumn{1}{c}{\textbf{Miss}} & \multicolumn{1}{c}{\textbf{FA}} & \multicolumn{1}{c}{\textbf{Error}}\\
    \midrule
    VAD1 & $1.38$ & $2.87$ & $4.26$ & $1.69$ & $3.24$ & $4.93$ \\
    VAD2 & $1.61$ & $2.6$ & $4.21$ & $1.96$ & $2.95$ & $4.91$ \\
    VAD3 & $1.38$ & $2.75$ & $4.13$ & $1.66$ & $3.09$ & $4.75$ \\
    pa & $1.17$ & $2.30$ & $3.48$ & $1.45$ & $2.88$ & $4.34$ \\
    pa\_tuned & $1.20$ & $2.26$ & $3.46$ & $1.49$ & $2.84$ & $4.33$ \\
    \bottomrule
  \end{tabular}
\end{table}

\begin{table}[th]
\footnotesize
  \caption{Hyper-parameters of pyannote models}
  \label{tab:vad2}
  \centering
    \begin{tabular}{ l c c c c }
    \toprule
    \multicolumn{1}{c}{\textbf{System}} & \multicolumn{1}{c}{\textbf{Onset}} & \multicolumn{1}{c}{\textbf{Offset}} & \multicolumn{1}{c}{\textbf{Min\_duration\_on}} & \multicolumn{1}{c}{\textbf{Min\_duration\_off}} \\
    \midrule
    pa & $0.767$ & $0.713$ & $0.182$ & $0.501$  \\
    pa\_tuned & $0.757$ & $0.727$ & $0.132$ & $0.481$ \\
    \bottomrule
  \end{tabular}
\end{table}

\subsection{Segmentation}

In system, we use the uniform segmentation method. \cite{park2021multi} proposed a method that can alleviate the trade-off restriction between temporal resolution and speaker fidelity. There is a complementary relationship between systems of different time scales. 

We found that in the case of a fixed window size, as the hop size decreases, DER can be improved to a certain extent. When the length of the window is different, the accuracy of the speaker representation is relatively more accurate for the longer segment of the window. Because the frame shift is the smallest unit to determine the speaker's identity, a small frame shifting is conducive to accurately locating the change point of the speaker.

\subsection{Speaker Embedding}
ECAPA-TDNN\cite{desplanques2020ecapa}, which achieved state-of-the-art performance on speaker recognition, was adopted to train the speaker embedding model. The training data consisted of voxceleb1\cite{Nagrani17} and voxceleb2\cite{Chung18b}. Pitch shifting and spectral augmentation \cite{park2019specaugment} are applied when training. The training process was implemented through speechbrain\cite{speechbrain}, an open-source toolkit for speech processing, which achieved an EER=0.59\% on voxceleb1 test set. In order to verify its performance on dev2, We merged the speech of each speaker in dev2 respectively according to the reference information, and cut them into sentences with a length of 1 to 5 seconds randomly. Then one hundred thousand trial pairs were formed from these sentences, in which the ratio of positive and negative pairs was guaranteed to be 1:1. As shown in Table~\ref{tab:eer}, dev2 reported a higher EER=3.40\% than voceleb1. As-norm\cite{matejka2017analysis} was adopted to improve performance and a better EER=3.17\% was obtained. However, as-norm did not bring similar improvements on DER when applied to speaker diarization, therefore, as-norm was not used in our speaker diarization system.

\begin{table}[th]
  \caption{EER on voxceleb1 test set and dev2.}
  \label{tab:eer}
  \centering
    \begin{tabular}{ c c c c }
    \toprule
    \multicolumn{1}{c}{\textbf{Test set}} & \multicolumn{1}{c}{\textbf{EER}} \\
    \midrule
    voceleb1  & $0.59\%$    \\
    dev2  & $3.40\%$    \\
    dev2+as-norm  & $3.17\%$    \\
    \bottomrule
  \end{tabular}
  
\end{table}

\subsection{Clustering}

\subsubsection{Initial Clustering}

We first perform NMESC\cite{park2019auto} on the audio segments. NMESC is a kind of spectral clutering algorithm, which uses normalized maximum eigengap (NME) values to estimate the number of clusters as well as the parameters for the threshold of the elements of each row in an affinity matrix, without parameter tuning on a development set. During NMESC, we found two tricks that can boost the DER of the final diarization system.

The first trick is called short duration filter, which is referred to the work of \cite{xiao2021microsoft}. 
The motivation of is to increase the chance of the main speakers being diarized correctly while sacrificing the performance on minor speakers. 
We filter out sequential label fragments shorter than a given duration on the clustering output label sequence. 
After that a SV step is performed to assign the most similar speaker label to these fragments. 
If the similarity between a cluster to be assigned and its most similar speaker cluster is lower than an SV threshold, the cluster will not be assigned to the speaker cluster but treated as unassigned cluster.
The minimum fragments duration and SV threshold are tuned on the development set and set to 2.5 seconds and 0.15 respectively.

We call the other trick under-clustering. The estimation of number of cluster is a step of spectral counting, corresponding to the estimating of number of speaker here.
We find that in dev1+dev2, 42.19\% of the estimated number of speakers is less than the ground truth.
We also find that the diarization DER will get lower if we provide the ground truth number of speaker in the clustering process.
So we manually enlarge the estimated number of speaker to make the speaker number a little "over-counted".
We tried some different methods and found that multiplying the estimated number of speaker by 1.2 achieved the lowest DER on development set, which is chosen for the following results.
Specifically, we found that this trick only works in system with time scale of 1.5s/0.75s.

\begin{table}[th]
  \caption{The influence of speaker counting on DER. SC refers to DER of NMESC. RC refers to DER of re-clutering. ground truth refers to use ground truth number of speaker instead of estimated ones in NMESC. estimation + n means add the estimated number of speaker by n if number of speaker greater than 3. estimation * n means multiplying the  number of speaker by n if number of speaker greater than 3 }
  \label{tab:sc1}
  \centering
    \begin{tabular}{ l c c c c }
    \toprule
    \multicolumn{1}{c}{\textbf{System}} & \multicolumn{2}{c}{\textbf{DER on dev1+dev2}} & \multicolumn{2}{c}{\textbf{DER on dev2}} \\
    \midrule
    \multicolumn{1}{c}{\textbf{}} & \multicolumn{1}{c}{\textbf{SC}} & \multicolumn{1}{c}{\textbf{RC}} &  \multicolumn{1}{c}{\textbf{SC}} & \multicolumn{1}{c}{\textbf{RC}}\\
    \midrule
    baseline & $6.58$ & $5.51$ & $7.31$ & $6.32$ \\
    ground truth & $12.67$ & $5.11$ & $14.26$ & $5.80$ \\
    estimation + 1 & $10.90$ & $5.46$ & $11.48$ & $6.30$ \\
    estimation + 2 & $15.29$ & $5.35$ & $16.34$ & $6.16$ \\
    estimation + 3 & $18.90$ & $5.34$ & $20.47$ & $6.16$ \\
    estimation * 1.2 & $12.12$ & $5.33$ & $12.89$ & $6.15$ \\
    estimation * 1.3 & $14.88$ & $5.36$ & $16.41$ & $6.18$ \\
    estimation * 1.5 & $19.82$ & $5.35$ & $21.73$ & $6.18$ \\
    \bottomrule
  \end{tabular}
\end{table}

\begin{table}[th]
\footnotesize
  \caption{OSD on dev2}
  \label{tab:overlap2}
  \centering
    \begin{tabular}{ l c c c c }
    \toprule
    \multicolumn{1}{c}{\textbf{Test set}} & \multicolumn{1}{c}{\textbf{Accuracy}} & \multicolumn{1}{c}{\textbf{Precision}} & \multicolumn{1}{c}{\textbf{Recall}} & \multicolumn{1}{c}{\textbf{F1 score}} \\
    \midrule
    dev2 & $97.20\%$ & $76.85\%$ & $36.81\%$ & $0.4978$  \\
    \bottomrule
  \end{tabular}
\end{table}

\subsubsection{Re-clustering}

\renewcommand{\multirowsetup}{\centering}
\begin{table*}[htbp]
        \centering
                \caption{The DER and JER of the proposed speaker diarization system.}
                \begin{tabular}{c|cccc|cc|cc|cc}
                        \toprule
                        \multirow{2}[0]{0.3in}{System} & \multirow{2}[0]{0.4in}{Time-scale} & \multirow{2}[0]{0.4in}{Under-cluster} & \multirow{2}[0]{0.4in}{SC-threshold} & \multirow{2}[0]{0.4in}{Re-cluster} & \multicolumn{2}{c|}{dev2} & \multicolumn{2}{c|}{dev1+dev2} & \multicolumn{2}{c}{test} \\
                        \cline{6-11}
                         & & & & & DER & JER & DER & JER & DER & JER \\
                        \hline
                        1 & 1.0s/0.25s & no  & 0.04 & 0.047       & 5.24\% & 28.38\% & 4.33\% & 23.80\% & -        & -       \\
                        2 & 1.0s/0.50s  & no  & 0.07 & 0.04        & 5.45\% & 30.44\% & 4.50\% & 25.71\% & -        & -       \\
                        3 & 1.5s/0.75s & yes & 0.10 & 0.048       & 5.48\% & 31.07\% & 4.51\% & 26.49\% & -        & -       \\
                        4 & \multicolumn{4}{c|}{fusion of 1,2,3}  & 5.17\% & 29.56\% & 4.27\% & 24.71\% & 5.15\% & 26.02\% \\
                        \bottomrule
                \end{tabular}%
                \label{tab:diarization_result}%
\end{table*}%

In the method proposed in this paper, the speaker representation is calculated on a short speech segment. The shorter the segment, the less robust the speaker’s representation will be. If the speaker's representation can be extracted in a longer segment, and then perform clustering with this, the system's result will be more robust.

"Re-clustering" is a step where all segments of a speaker in the recording are concatenated to extract a new embedding. Then, use a cluster method to join speakers who are similar. It is proposed by BUT \cite{landini2021analysis}.

In the clustering process in the previous step, the number of speakers will be estimated. According to our statistics, the system 3 in Table \ref{tab:diarization_result}, only 50.90\% of the recordings’ speakers are correctly estimated. There are 42.19\% recordings that are lower than the actual number of speakers, and 6.92\% recordings are higher than the actual number of speakers.

When the number of speakers is lower than the actual number of speakers, using re-clustering may result in more speaker error rates. But when the number of speakers is higher than the actual number of speakers, this method can effectively reduce the speaker error rate. In order to optimize DER result, the parameters of re-clustering need to be adjusted reasonably. In development data set in this challenge, the best results can be obtained by using parameters around 0.047.

\subsection{Overlapped speech detection and handling}

Overlap speech detection (OSD) is task of estimating regions where two or more speakers are speaking simultaneously. To obtain a satisfactory DER, OSD is an essential component, without which, even a perfect diarization system, would misses the second or more speakers in overlap regions, leading to an increase in DER according to the duration of overlap regions. 

A blind source separation system was first used in OSD, in which we assumed that there are at most two speakers in each record. Two channel outputs are obtained from the separation model, and the intersection of their VAD outputs is regarded as the overlap area. However, experiments on dev1 and dev2 showed that the performance of the separation model is unsatisfied, appearing that the voice of the same speaker appeared in both channels. The model in pyannote\cite{bredin2021}, trained on a compound training set for DIHARD3\cite{ryant2020third}, AMI\cite{carletta2007unleashing} and dev1, was chosen to perform OSD. Grid search was performed on the four thresholds to obtain the best DER of dev2, the corresponding precision and recall rate of which were 76.85\% and 36.81\% respectively, as shown in Table~\ref{tab:overlap2}.

According to the analysis of the data set, the overlapping fragments in the development set accounted for 3.28\%, while the overlapping fragments of three or more people accounted for only 0.6\%. The proportion of overlapping fragments of three or more speakers is quite small. So we only discuss the case where the overlapping segment includes two human voices. Assuming that only two people in the overlapping segment are talking at the same time, the speaker should be the closest speaker in the time series before and after the overlapping segment. Based on this assumption, we constructed our overlap handling method.

\subsection{System fusion}

To make the diarization outputs more robust, we adopted a multi-system fusion method. Recently, \cite{raj2021dover} proposed a DOVER-Lap algorithm, which combines the different system hypotheses with overlap-aware.  It uses an approximation algorithm that is globally informed by all pair-wise costs, and it can assign multiple speaker labels to a region. This method can be used not only for the fusion of multi-systems, but also for the later fusion of multi-channel system.

We believe that systems of different time scales have a certain complementary relationship. Therefore, in the final submission system, we combined three hypotheses with different time scales. We use the best system at each time scale as the fusion subsystem. The three subsystems correspond to 0.25s frame shift, 0.5 frame shift and 0.75s frame shift during speaker feature extraction, and the corresponding voting weights are 0.4, 0.3, and 0.3 respectively.

\section{Results and analysis}

The diarization result of the proposed systems are shown in Table \ref{tab:diarization_result}. The diffrence between systems 1-3 is time scale, auto-tuning spectral clustering's rp threshold, and the re-clustering parameter. System 3 then introduced a under clustering method into the auto-tuning spectral clustering. On a single system, System 1 obtains the best DER 4.33\% and best JER 23.80\%.

Finally, systems 1, 2 and 3 were fused using the DOVER-Lap algorithm. Compared with the best single system 1, DER has been improved by 0.05\%, but the JER has been reduced by 0.91\%. Although the system fusion result on DER has improved, the reasons for the decline in JER result need to be further analyzed. Our best submissions on the evaluation set are DER 5.15\% and JER 16.02\%.

\section{Conclusions}

This paper describes the ByteDance speaker diarization system, evaluated at the fourth track of the VoxCeleb Speaker Recognition Challenge 2021. We proposed the diarization system consists of VAD, ECAPA-TDNN-based speaker embedding extractor, auto-tunning spectral clustering with short duration filter, AHC-based recluster, overlap handle and DOVER-Lap for system fusion. The final system achieves DER 4.27\% and DER 5.15\% on development and evaluation set, respectively, being ranked second at the VoxSRC 2021.

\bibliographystyle{IEEEtran}

\bibliography{mybib}
\end{document}